# Anti-electrostatic hydrogen bonding between anions of ionic liquids：A density functional theory study

Junwu Chen,[1,2] Kun Dong,[1] Lei Liu,[1]* Xiangping Zhang,[1,2] and Suojiang Zhang[1,2]*

[1] Beijing Key Laboratory of Ionic Liquids Clean Process, CAS Key Laboratory of Green Process and Engineering, State Key Laboratory of Multiphase Complex Systems, Institute of Process Engineering, Chinese Academy of Sciences, Beijing 100190, P.R. China

[2] School of Chemical Engineering, University of Chinese Academy of Sciences, Beijing 100049, P.R. China

Corresponding authors: liulei@ipe.ac.cn; sjzhang@ipe.ac.cn

# ABSTRACT

Hydrogen bonds (HBs) play a crucial role in the physicochemical properties of ionic liquids (ILs). At present, HBs between cations and anions (Ca-An) or between cations (Ca-Ca) in ILs have been reported extensively. Here, we provided DFT evidences for the exists of HBs between anions (An-An) in the IL 1-(2-hydroxyethyl)-3-methylimidazolium 4-(2-hydroxyethyl)imidazolide [HEMIm][HEIm]. The thermodynamics stabilities of anionic, cationic, and $H_2O$ dimers together with ionic pairs were studied by potential energy scans. The results show that the cation-anion pair is the most stable one, while the HB in anionic dimer possesses similar thermodynamics stability to the water dimer. The further geometric, spectral and electronic structure analyses demonstrate that the inter-anionic HB meets the general theoretical criteria of traditional HBs. The strength order of four HBs in complexes is cation-anion pair > $H_2O$ dimer ≈ cationic dimer > anionic dimer. The energy decomposition analysis indicates that induction and dispersion interactions are the crucial factors to overcome strong Coulomb repulsions, forming inter-anionic HBs. Lastly, the presence of inter-anionic HBs in ionic cluster has been confirmed by a global minimum search for a system containing two ionic pairs. Even though hydroxyl-functionalized cations are more likely to form HBs with anions, there still have inter-anionic HBs between hydroxyl groups in the low-lying structures. Our studies broaden the understanding of HBs in ionic liquids and support the recently proposed concept of anti-electrostatic HBs.

**Keywords:** Ionic liquids, Hydrogen bonds, Ionic pairs, Density functional theory

## I. INTRODUCTION

The nature of hydrogen bonds (HBs) has attracted long-lasting attention since its discovery in 1920,[1-4] due to their significance in chemistry, biology and material science.[5, 6] By the 100th anniversary,[7] several new types of HBs have been reported, e.g. blue-shifted HBs, dihydrogen HBs, inverse HBs, and ionic HBs.[5] From the theoretical point of view, the electrostatic attractions have been considered as the essential character of HBs.[3, 4] The HBs between opposite-charged ions have been demonstrated showing different features due to strong Coulomb attractions.[4, 8-11] Recently, several theoretical and experimental evidences for HBs between the same charged ions have been reported.[12-16] These counter-intuitive phenomena which challenge the classic concept of HBs are gradually accepted by researchers and prompted IUPAC to revise the definition of HBs in 2011.[4, 9, 17] In 2014, Weinhold et al.[4] proposed the concept of anti-electrostatic HBs and classified the HBs between the same charged ions as the new class of HBs. They characterized some HBs in anion-anion or cation-cation complexes and investigated their covalent features.[4]

Ionic liquids (ILs), comprised of anions and cations, show many advantages such as low vapor pressure and wide electrochemical window.[5, 6, 18-20] The HB networks in ILs play a very important role in their physicochemical properties.[5, 6, 21-23] Gamrad et al.[24] first demonstrated the existence of HBs between cations in the crystals of ILs by X-ray diffraction analysis. In 2015, Ludwig et al. provided the first spectroscopic evidence for HBs between cations, and DFT calculations well supported the results of experiments.[25-27] Advanced characterization techniques, such as cryogenic ion vibrational predissociation spectroscopy and neutron diffraction, have also been employed to further investigate the inter-cationic HBs in ILs.[28-30] Those studies suggested several important factors for the formation of HBs between cations (such as the basicity of anions) and proved that inter-cationic HBs could be stronger than those between opposite-charged ions through cooperative effects.[28-32] Moreover, some studies demonstrated that this new class of HBs could give ILs unusual physicochemical properties, i.e. increasing their melting point and viscosity, reducing their conductivity, inhibiting crystallization under supercooled conditions and changing the interfacial nanostructures of ILs.[25, 33-35] However, to the best of our knowledge, no research on the HBs between the anions of ILs has been reported so far.

Here, we selected the IL 1-(2-hydroxyethyl)-3-methylimidazolium 4-(2-hydroxyethyl)imidazolide [HEMIm][HEIm] (see in Fig. 1) to explore possible HBs between anions (hydroxyl groups, OH). The imidazolide anion with hydroxyethyl tether was chosen as the anion. The presence of the hydroxyethyl tether could increase the distance between the OH and the negative charge center, thereby weakening the Coulomb repulsions during the formation of HBs between the anions.[26, 30] Besides, organic anions have better diversity and more excellent tunability than the inorganic anions, which facilitates a more flexible design of ILs that may contain HBs between anions. Imidazolium cation, one of the most common cations in ILs, was selected as the cation.[36] The thermodynamic stabilities of anionic dimers, cationic dimers, $H_2O$ dimers and ionic pair were explored by the potential energy scan. The reasonable energy well-depth of anionic dimer indicates that inter-anionic HBs could overcome strong electrostatic repulsions and possess appreciable thermodynamic stability. The HBs were further analyzed by atoms in molecules (AIM), natural bond orbital (NBO), as well as the energy decomposition analysis. Lastly, the global optimization search was performed to confirm the presence of HB between anion in a system consisting of two ionic pairs.

## II. COMPUTATIONAL METHODS

Geometry optimization and harmonic frequency calculations for all of the structures were performed using the Gaussian 09 (revision D.01) package[37] at dispersion-corrected B3LYP-D3/6-31+G* level. The positive frequencies ensure that all structures are local minima. Topological analysis of electronic density (Quantum Theory of Atoms In Molecules, QTAIM) and reduced density gradient (RDG) analysis were calculated based on the B3LYP-D3/6-31+G* wave function at the optimized geometries using the Multiwfn version 3.6 code.[38] Natural bond orbital (NBO) analyses have been carried out using the NBO (version 3.1) software[39] which is a component of Gaussian 09 program. The analyses of interaction energy based on symmetry adapted perturbation theory (SAPT) were performed at the SAPT2+/aug-cc-pVDZ level on the optimized geometries. The PSI4 code[40] was employed for the SAPT analyses. To find the most stable structure of the ionic clusters, 500 random initial configurations were generated by Molclus program[41] and preliminarily optimized by a semi-empirical method at the PM7 level using MOPAC2016 program.[42] Then, 100 low-lying structures were chosen and optimized at the B3LYP-D3/6-31+G* level.

## III. RESULTS AND DISCUSSION

The potential energy curves of the anionic dimer (HEIm$^-$)$_2$ were calculated by three methods (B3LYP, B3LYP-D3, MP2) with well-balanced 6-31+G* basis set.[8, 30, 31] The high-precision MP2 method is used to benchmark the accuracy of B3LYP and B3LYP-D3 methods. As shown in Fig. S1, all three potential energy curves have shallow energy wells, in which, the B3LYP-D3 method has a similar well-depth (1.3 kcalmol$^{-1}$) to that of MP2 method (1.4 kcalmol$^{-1}$), while B3LYP method has a smaller well-depth (0.6 kcalmol$^{-1}$). Besides, the OH⋯OH distance of local minimum calculated by B3LYP-D3 method (1.97 Å) is closer to MP2 method (1.98 Å) than that of B3LYP method (2.08 Å) (see in Table S1). Thus, B3LYP-D3 was employed to describe geometric and energetic properties in next discussions.

The relaxed potential energy curves of ionic pair, H$_2$O dimer, cationic dimer and anionic dimer were calculated (see Fig. 2). The ionic pair has the deepest energy well (-102.4 kcalmol$^{-1}$) at $R_{H⋯O}$ = 1.86 Å and the well-width is relatively large. The deep and wide energy trap can effectively prevent tunneling-type dissociation and Coulomb explosion.[9, 30] The H$_2$O dimer possesses a wide energy well with a depth of -7.2 kcalmol$^{-1}$ (much smaller than the ionic pair) at $R_{H⋯O}$ = 1.90 Å, which is similar to the reported results (7.5 kcalmol$^{-1}$, $R_{H⋯O}$ = 1.9 Å).[43] Due to strong electrostatic attractions, the ionic pair is significantly more stable than H$_2$O dimer which are linked by traditional HBs. In contrast to the ionic pair and H$_2$O dimer, charged ionic dimers have to overcome the strong long-range electrostatic repulsions through the quantum short-range HBs.[4, 9] As shown in Fig. 2, the cationic dimer has an energy well of 2.4 kcalmol$^{-1}$ at $R_{H \cdots O}$ = 1.88 Å, which are similar to the reported results (0.7 kcalmol$^{-1}$, $R_{H \cdots O}$ = 2.00 Å).[30] The anionic dimer has the smallest energy well of 1.3 kcalmol$^{-1}$ at $R_{H \cdots O}$ = 1.97 Å. Due to the electrostatic repulsion, well depths of both cationic and anionic dimers are smaller than that of the H$_2$O dimer. However, they all have the same order of magnitude, indicating that the HBs in cationic and anionic dimers possess appreciable thermodynamic stability compared to traditional neutral HBs. Moreover, the larger well depth and shorter equilibrium distance suggest that HB in cationic dimer is more stable than that in anionic dimer. The cationic character can lower the energy and enhance the polarity of the antibonding orbital ($\sigma^*_{O-H}$), while the anionic character can raise the energy and enhance the diffuseness of the lone-pair electron orbital (O$_{lp}$).[5] Thus, the inter-cationic HBs are more robust than the inter-anionic HBs due to the reinforced polarization, which is consistent with the reported literature.[5]

The influence of the hydroxyalkyl length on the thermodynamic stability of anionic dimers was further investigated by the potential energy scan. As shown in Fig. 3, the energy well-depth increases substantially due to elongation of the hydroxyalkyl chain. For hydroxybutyl chain, the energy well-depth of the anionic dimer is 3.2 kcalmol$^{-1}$, which is twice more than that of the hydroxyethyl chain (1.3 kcalmol$^{-1}$). The longer hydroxyalkyl chain increases the distance between the negative charge centers of anions, resulting in weaker electrostatic repulsion and higher thermodynamic stability.[26, 30] The equilibrium HB lengths are also shortened from 1.97 Å to 1.93 Å with longer hydroxyalkyl chain. The reduction of the OH$\cdots$O distance can promote the orbital overlap ($O_{lp} \rightarrow \sigma^*_{O-H}$) and enhance covalent components, thereby strengthening inter-anionic HBs.[5]

As shown in Table I, the local minimum structures of ionic pair, H$_2$O, cationic and anionic dimers have typical HB lengths of 1.86 Å, 1.90 Å, 1.88 Å and 1.97 Å with bond angles of 175.7°, 171.1°, 162.9° and 171.6°, respectively. The ionic pair has the shortest HB length (1.86 Å), while anionic dimer has the longest HB length (1.97 Å). The rational OH···OH distances and near-linear O-H···O bond angles can be considered as the characteristics of HBs.[4] The infrared spectra of local minimums were also simulated. As shown in Table I and Fig. S2, the −OH stretching vibration bands of dimers and ionic pair exhibit significant redshifts (107 cm$^{-1}$, 112 cm$^{-1}$, 167 cm$^{-1}$, 276 cm$^{-1}$), which are attributed to the formation of HBs between the OH groups.[8, 25, 26] The charge transfer from the lone pair of O ($n_O$) into the antibonding orbital of OH ($\sigma^*_{OH}$) weakens the O−H bond (i.e. bond length becomes longer) and decrease its stretching frequency.[44-46] The OH vibration band of ionic pair has the largest displacement (276 cm$^{-1}$) to lower wavenumber, which implies that it has a stronger HB than other dimers. Overall, the data of geometric parameters and harmonic frequencies well support the formation of HBs in anionic and cationic dimers. To further investigate the electronic properties of HBs in dimers and the ionic pair, we performed AIM, NBO and RDG analyses based on the local minimum structures.

AIM molecular graphs and NBO 3$d$ diagrams of the $n_O \rightarrow \sigma^*_{OH}$ overlaps are shown in Fig. S3. The OH···O interactions in four complexes exhibit typical AIM-type bond paths and (3, -1) bond critical points (BCPs), which conforms to the characteristics of

HBs defined by IUPAC.[5, 17] The related data of the electron density ($\rho_{BCP}$), Laplacian of the electron density ($\nabla^2\rho_{BCP}$), the total electron energy density ($H_{BCP}$) and second-order interaction energy ($E^{(2)}$) are listed in Table II. The positive values (0.099, 0.092, 0.096, 0.078) of $\nabla^2\rho_{BCP}$ indicate that the HBs in four complexes are non-covalent bonds, while the negative values (-0.0015, -0.0009, -0.0008, -0.0009) of $H_{BCP}$ imply that these HBs possess covalent properties.[5, 47] The large $E^{(2)}$ values (16.3, 12.9, 12.6, 9.5 kcalmol$^{-1}$) also indicate that the HBs in the four complexes have covalent components.[4, 5] The donor-acceptor covalent force could contribute to overcoming the strong long-range Coulomb repulsions.[4, 31] Moreover, there is no obvious difference in the results of QTAIM and NBO analyses of inter-ionic HBs (ionic pair, anionic dimer, and cationic dimer) compared to traditional neutral HBs ($H_2O$ dimer). The HBs in each complex meet all commonly accepted electronic structure criteria of HBs[4, 5], in particular, these of anionic dimer and cationic dimer which have strong electrostatic repulsions. According to the classification criteria of HBs [5], all four HBs of the complexes belong to medium-strength HBs. The strength sequence of four HBs is ionic pair > $H_2O$ dimer ≈ cationic dimer > anionic dimer, which is somehow consistent with well-depth of potential energy curves depicted in Fig. 2.

Subsequently, RDG analysis was used to study the weak interactions in four complexes. As depicted in Fig. 4a, the leftmost spike (-0.033 a.u.) in the scatter diagram corresponds to the HB between hydroxyl groups (the blue circular RDG isosurface in the surface diagram).[48, 49] The multiple spikes in the sign($\lambda_2$)$\rho$ range of -0.02 to 0.02 are assigned to C-H···π, C-H···σ and van der Waals interactions.[47-49] The most negative

sign($\lambda_2$)$\rho$ value of HB (OH···O) indicates that HB was the strongest attraction between the cation and anion. As for H$_2$O dimer (Fig. 4b), only one spike with negative sign($\lambda_2$)$\rho$ value (-0.029) is found in the scatter diagram, which is attributed to HB (OH···O) interaction. For cationic and anionic dimers containing strong electrostatic repulsion, blue spikes (-0.030 a.u., -0.025 a.u.) and blue circular RDG isosurfaces between hydroxyl groups are also found in the scatter and surface diagrams, respectively. It further demonstrates that HBs can be formed between the hydroxyl groups of the same-charged ions. The more negative sign($\lambda_2$)$\rho$ and the deeper blue isosurface of HB (OH···O) also prove that the HB between anion and cation is stronger than inter-cationic HB and inter-anionic HB.[48, 49]

Next, energy decomposition calculations were performed to reveal the dominating component in HBs interactions (see in Fig. 5 and Table S2). For the ionic pair, electrostatic attraction (-110.6 kcalmol$^{-1}$) is the predominant factor in the formation of HB, while the contributions of induction and dispersion (-25.0 kcalmol$^{-1}$, -21.2 kcalmol$^{-1}$) are relatively small. The large negative total interaction energy (-108.2 kcalmol$^{-1}$) of the ionic pair indicates that the ionic pair linked by HB is thermodynamically stable and is easy to form. The H$_2$O dimer possesses a negative but small total interaction energy (-4.7 kcalmol$^{-1}$). Similar to the ionic pair, electrostatics (-9.2 kcalmol$^{-1}$) is the main attractive component of HB in H$_2$O dimer and the contributions of induction and dispersion are relatively small, being -2.9 kcalmol$^{-1}$, -2.4 kcalmol$^{-1}$, respectively. However, cationic dimers and anionic dimers have positive total interaction energies, indicating that the formation of cationic and anionic dimers have to overcome certain amount of energy barriers. In contrast to the ionic pair and H₂O dimer, the formation of HBs in cationic and anionic dimers is derived from induction and dispersion, while the electrostatics term (+23.0 kcalmol$^{-1}$, +21.7 kcalmol$^{-1}$) becomes the dominant component of repulsion. The results indicate that electrostatic attraction is an important but not critical factor in the formation of HBs, which is consistent with the literature.[4, 8, 9, 24] Moreover, in the presence of strong electrostatic repulsion, HBs may still form between anions duo to induction and dispersion force.

Lastly, we performed global optimization calculations of ionic clusters containing two pairs of anions and cations to confirm the presence of HBs between anions, and found 44 low-lying structures with relative energy differences less than 9.5 kcalmol$^{-1}$ (see Fig. S4-6). Here, we mainly focus on the HBs between the OH group in the ionic clusters, and the low-lying structures could be grouped into three categories: (1) containing HBs between the OH of cations and anions (Ca–OH⋯OH–An), (2) containing HBs between the OH group of anions (An–OH⋯OH–An), and (3) no HBs between OH groups. For each category, the most stable structures have been selected for detail discussion, which are provided in Fig. 6. The lowest-energy structure (Fig. 6, I) has seven HBs but none of them is between OH groups. There are four OH groups in the lowest-energy structure, three of which form HBs with the nitrogen atoms of anions, and one of them forms HB with the C$^2$–H on the cation. Ca–OH⋯OH–An HBs are observed in 17 low-lying structures and first appear in the structure II (center structure in Fig. 6). Except for the OH groups in Ca–OH⋯OH–An, the other two OH groups in the structure II form HBs with the nitrogen atoms on the anion (Ca–OH⋯N–An and

An–OH⋯N–An). The energy difference between the structure II and the lowest-energy structure is rather small (only 1.3 kcalmol$^{-1}$), indicating that Ca–OH⋯OH–An HBs have a high probability of appearing in the studied ionic clusters. Importantly, An–OH⋯OH–An HBs are also observed in five low-lying structures and first appear in the structure III (right structure in Fig. 6). Except for the OH groups in An–OH⋯OH–An, the other two OH groups in the structure III form HBs with the nitrogen atoms on the anion (Ca–OH⋯N–An). The energy difference between the structure III and the lowest-energy structure is 4.7 kcalmol$^{-1}$, indicating that An–OH⋯OH–An HBs could also appear in ionic clusters. Surprisingly, there are no HBs between the OH groups of cations (Ca-OH⋯OH–Ca) in all 44 low-lying structures. It may be due to the interference of the negatively charged nitrogen atoms on the anions. The Ca–OH⋯N–An HBs are observed in all 44 low-lying structures, while Ca–OH⋯OH–An appear in only 17 low-lying structures. The HBs between anions are observed in 28 low-lying structures. Among them, 22 low-lying structures contain An–OH⋯N–An HBs, while only 5 low-lying structures contain An–OH⋯OH–An HBs. The results show that the nitrogen atoms (strong HB acceptor) on the anions could hinder the formation of HBs between the OH groups. In case the possibility of An–OH⋯OH–An is expected to be increased, the HB acceptor ability of the nitrogen atom of anion needs to be weakened (e.g., via enhancing steric effects). Besides, intra-anionic HBs (between OH group and N atom of one anion) are observed in 15 low-lying structures (e.g., structure 2 in Fig. S4), which could prevent the formation of An–OH⋯OH–An HBs. Based on the above-mentioned geometric parameter analysis, we proposed three important factors for the formation of inter-anionic HBs between OH groups in ILs: (1) larger distances between the negative charge center and the OH group of the anion, (2) enhancing the HB donor/acceptor ability of OH groups and weakening the HB acceptor ability of other sites on the anion, and (3) inhibiting the formation of intra-anionic HB (such as increasing the distance between the HB donor site and acceptor site in one anion).

## IV. CONCLUSION

We performed DFT calculations to predict the existence of HBs between anions in ILs. The relaxed potential energy scans show that the ionic pair has the largest energy well-depth of 102.4 kcalmol$^{-1}$ while cationic and anionic dimers have well-depths of 2.4 and 1.3 kcalmol$^{-1}$ (slightly smaller than that of the water dimer, 7.2 kcalmol$^{-1}$), respectively, and the strength sequence of HBs is found to be ionic pair > H$_2$O dimer ≈ cationic dimer > anionic dimer. The electronic calculations (i.e. QTAIM and NBO) show that the HBs in anionic dimer has appreciable thermodynamic stability and complies with typical theoretical criteria for HBs. Dispersion and induction forces are the key factors to overcome the strong Coulomb electrostatic repulsion between anions during the formation of HBs. For ionic clusters (HEMIm$^+$)$_2$(HEIm$^-$)$_2$, the global minimum search identifies five low-lying structures contain An–OH···OH–An HBs. The presence of cations could help anions overcoming electrostatic repulsion. However, strong HB acceptor sites of anions and intra-anionic HBs could hinder the formation of inter-anionic HBs between OH groups.

**SUPPLEMENTARY MATERIAL**

The potential energy curves at different theoretical levels; simulated infrared spectra, NBO 3d overlap diagrams, QTAIM molecular graphs, all low-lying structures of global optimization search, and some supporting results by calculation.

**ACKNOWLEDGMENTS**

This work was financially supported by the National Natural Science Foundation of China (21978294, 21838010, 21921005), and the CAS Pioneer Hundred Talents Program (L. L).

**DATA AVAILABILITY**

The data that support the findings of this study are available within the article or supplementary material and from the corresponding author upon reasonable request.

**TABLE LEGENDS**

**TABLE I.** The geometric parameters and −OH stretching frequencies for the HBs in dimers and ionic pair.

**TABLE II.** Results of QTAIM and NBO of HBs in the dimers and ionic pairs.

**TABLES**

**TABLE I.** The geometric parameters and −OH stretching frequencies for the HBs in dimers and ionic pair.

|  | HB length (Å) | Bond angle (°) | $v(OH)$ (cm$^{-1}$) | Redshift of $v(OH)$ (cm$^{-1}$) |
|---|---|---|---|---|
| Cation-Anion | 1.86 | 175.7 | 3485 | 276 |
| H$_2$O-H$_2$O | 1.90 | 171.1 | 3630 | 107 |
| Cation-Cation | 1.88 | 162.9 | 3594 | 167 |
| Anion-Anion | 1.97 | 171.6 | 3621 | 112 |

**TABLE II.** Results of QTAIM and NBO of HBs in the dimers and ionic pairs.

|  | $\rho_{BCP}$ (a.u.) | $\nabla^2\rho_{BCP}$ (a.u.) | $H_{BCP}$ (a.u.) | $E^{(2)}$ (kcalmol$^{-1}$) |
|---|---|---|---|---|
| Cation-Anion | 0.033 | 0.099 | -0.0015 | 16.3 |
| H$_2$O-H$_2$O | 0.029 | 0.092 | -0.0009 | 12.9 |
| Cation-Cation | 0.030 | 0.096 | -0.0008 | 12.6 |
| Anion-Anion | 0.025 | 0.078 | -0.0009 | 9.5 |

**FIGURE LEGENDS**

**FIG. 1**. Investigated HEMIm$^+$ cation and HEIm$^-$ anion.

**FIG. 2**. Relaxed-scan potential energy curves of dimers and ionic pair at B3LYP-D3/6-31+G* level, and corresponding minimum structures. Color legend: C grey, H white, O red, N blue.

**FIG. 3**. Relaxed-scan potential energy curves for the anionic dimers with different hydroxyalkyl lengths at B3LYP-D3/6-31+G* level.

**FIG. 4**. The RDG scatter diagrams (left) and surface diagrams (right) of ionic pair (a), H$_2$O dimer (b), cationic dimer (c) and anionic dimer (d).

**FIG. 5**. Interaction energy decomposition results for dimers and the ionic pair at SAPT2+/aug-cc-pVDZ level.

**FIG. 6**. The lowest-energy structure (I), the most stable low-lying structures containing Ca–OH···OH–An (II) or An–OH···OH–An (III) for ionic clusters (HEMIm$^+$)$_2$(HEIm$^-$)$_2$ at B3LYP-D3/6-31+G* level. The values in parentheses are the energy differences (kcalmol$^{-1}$) from the lowest-energy structure. Color legend: C grey, H white, O red, N blue. The H atoms that do not participate in HBs are omitted for clarity. The red dashed lines represent the HBs between OH groups and the black dashed lines represent other HBs (too weak HBs are ignored).

**FIGURES**

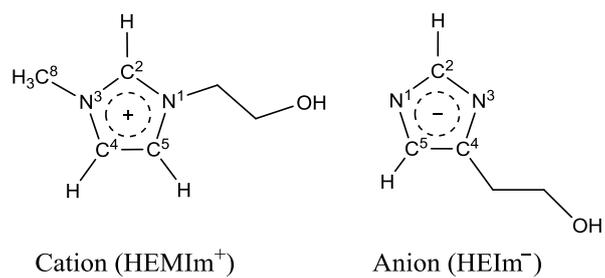

Cation (HEMIm$^+$)    Anion (HEIm$^-$)

**FIG. 1**. Investigated HEMIm$^+$ cation and HEIm$^-$ anion.

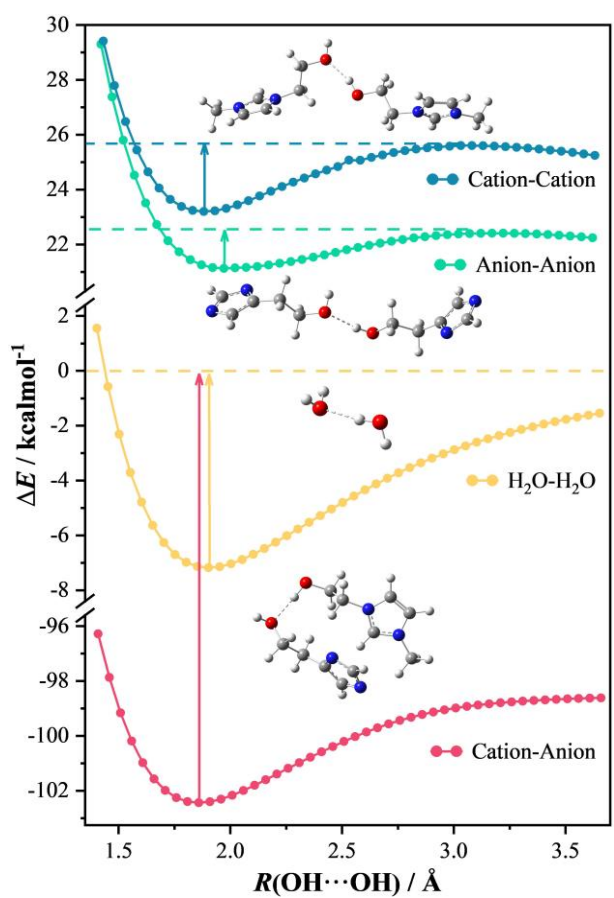

**FIG. 2**. Relaxed-scan potential energy curves of dimers and ionic pair at B3LYP-D3/6-31+G* level, and corresponding minimum structures. Color legend: C grey, H white, O red, N blue.

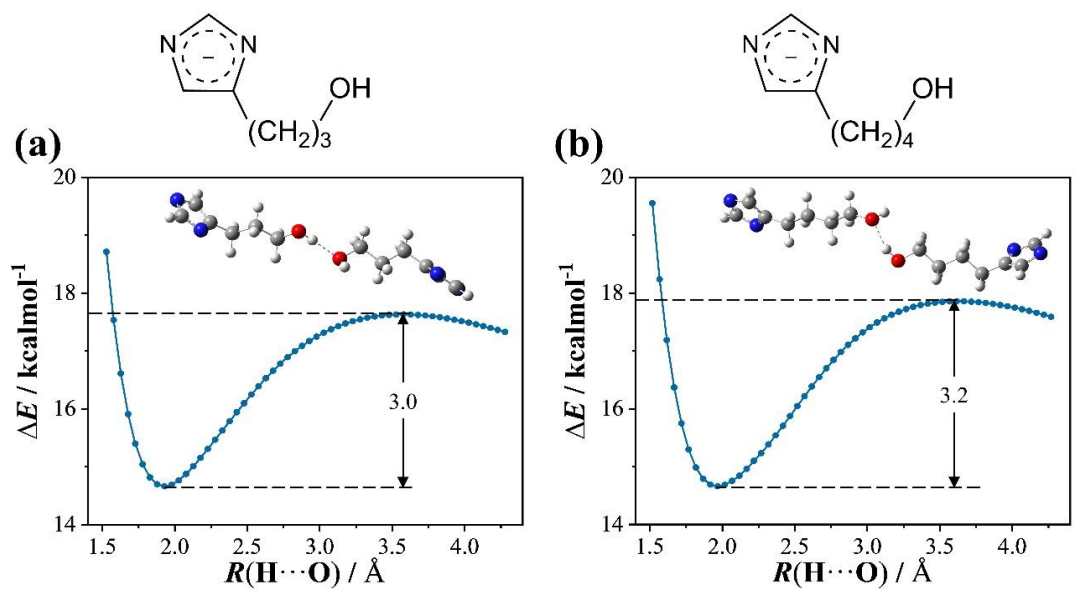

**FIG. 3**. Relaxed-scan potential energy curves for the anionic dimers with different hydroxyalkyl lengths at B3LYP-D3/6-31+G* level.

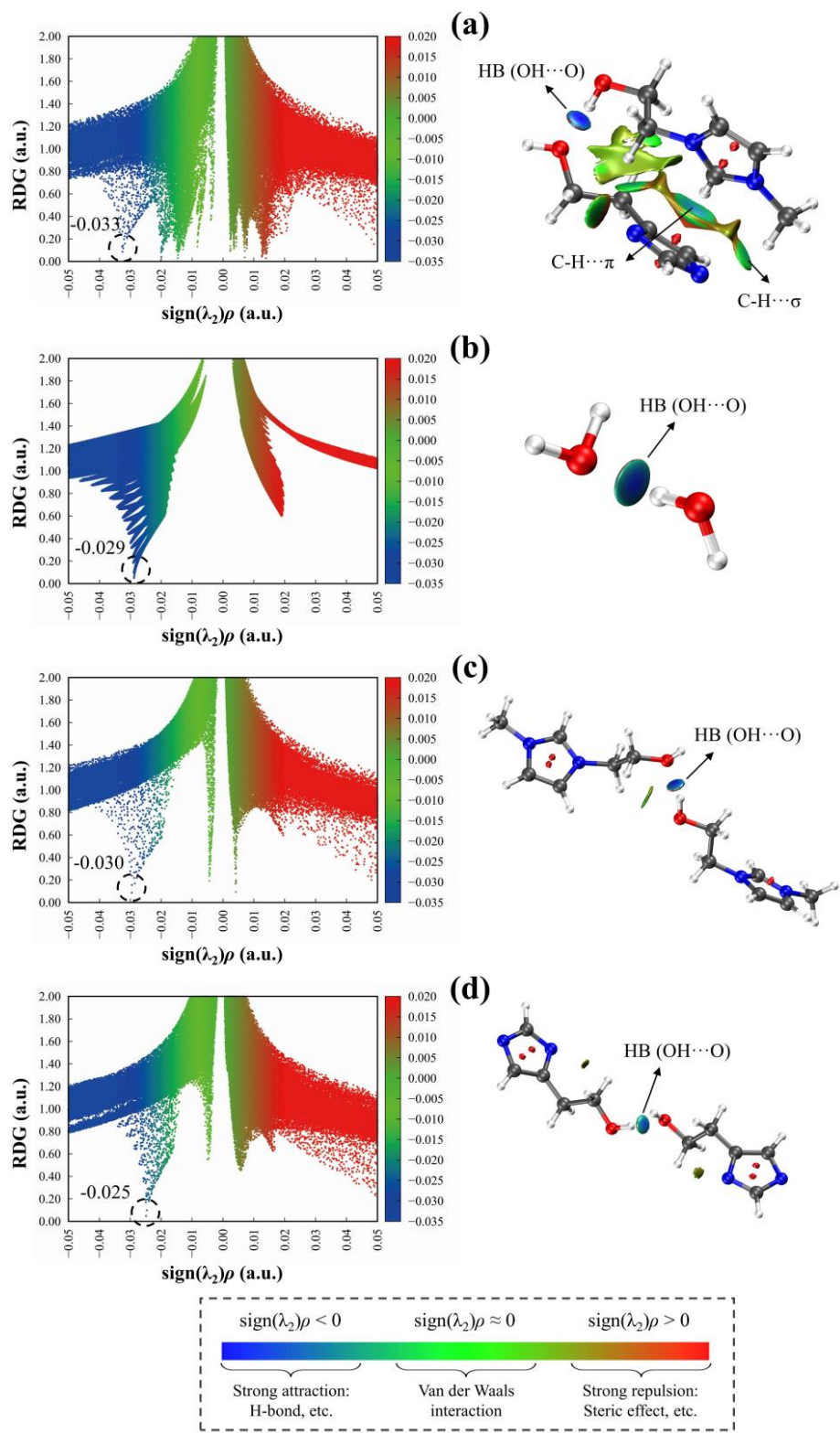

**FIG. 4**. The RDG scatter diagrams (left) and surface diagrams (right) of ionic pair (a), H$_2$O dimer (b), cationic dimer (c) and anionic dimer (d).

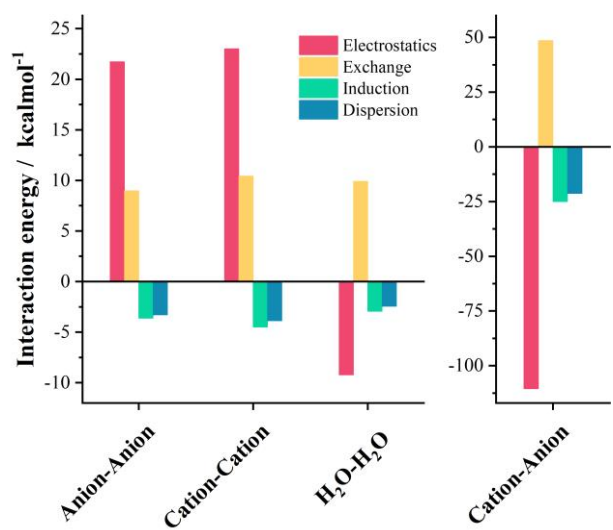

**FIG. 5**. Interaction energy decomposition results for dimers and the ionic pair at SAPT2+/aug-cc-pVDZ level.

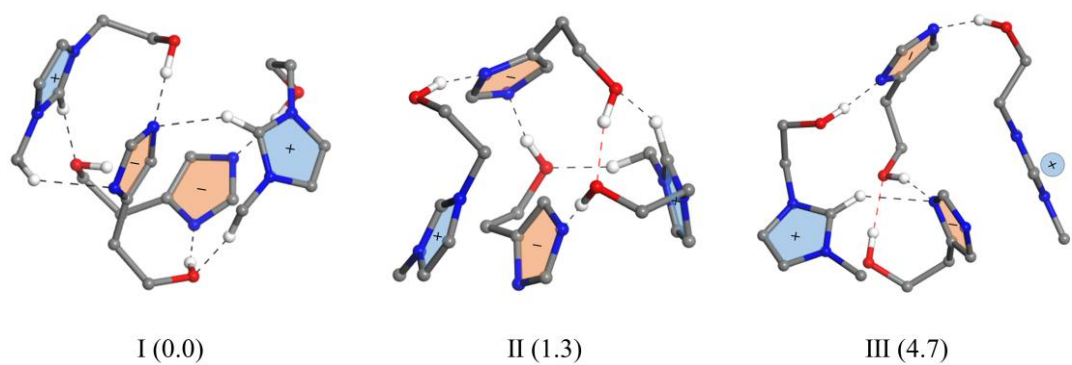

I (0.0)  II (1.3)  III (4.7)

**FIG. 6**. The lowest-energy structure (I), the most stable low-lying structures containing Ca–OH···OH–An (II) or An–OH···OH–An (III) for ionic clusters $(HEMIm^+)_2(HEIm^-)_2$ at B3LYP-D3/6-31+G* level. The values in parentheses are the energy differences (kcalmol$^{-1}$) from the lowest-energy structure. Color legend: C grey, H white, O red, N blue. The H atoms that do not participate in HBs are omitted for clarity. The red dashed lines represent the HBs between OH groups and the black dashed lines represent other HBs (too weak HBs are ignored).